\documentclass{article}

\usepackage{latexsym}

\newtheorem{theorem}{Theorem}[section]
\newtheorem{lemma}[theorem]{Lemma}

\newtheorem{definition}[theorem]{Definition}

\newcommand{\qed}{\;\;\;\Box} \newenvironment{proof}{\par{\bf
Proof:}}{\(\qed\) \par}

\begin{document}

\title{Deterministic approximation for the cover time of trees}

\author{Uriel Feige\thanks{Department of Computer
Science and Applied Mathematics, Weizmann Institute, Rehovot
76100, Israel. {\tt uriel.feige@weizmann.ac.il}.}\  \and Ofer
Zeitouni\thanks{Department of Mathematics, Weizmann Institute,
Rehovot 76100, Israel, and School of Mathematics, University of
Minnesota. {\tt ofer.zeitouni@weizmann.ac.il}.}}

\maketitle

\begin{abstract}
We present a deterministic algorithm that given a tree $T$ with
$n$ vertices, a starting vertex $v$ and a slackness parameter
$\epsilon
> 0$, estimates within an additive error of $\epsilon$ the {\em
cover and return time}, namely, the expected time it takes a
simple random walk that starts at $v$ to visit all vertices of $T$
and return to $v$. The running time of our algorithm is polynomial
in $n/\epsilon$, and hence remains polynomial in $n$ also for
$\epsilon = 1/n^{O(1)}$. We also show how the algorithm can be
extended to estimate the expected cover (without return) time on trees.
\end{abstract}

\section{Introduction}

Let $G$ be a connected graph with vertices $v_1, \ldots v_n$. We
consider simple random walks on $G$. Namely, the walk starts at
some vertex of the graph, and at every time step picks at random
with uniform probability a neighbor of the current vertex and
moves to it. Let $C_1^+(G)$ (the {\em expected cover and return
time}) denote the expected number of steps it takes a random walk
that starts at $v_1$ to visit all vertices of $G$ and return to
$v_1$. An empirical
estimate for the value of $C_1^+(G)$ can be
obtained by starting a random walk at $v_1$ and counting the
number of steps until it visits all vertices of $G$ and returns to
$v_1$. Averaging multiple such
estimates one obtains with high
probability an accurate approximation for $C_1^+(G)$. An
approximation within a multiplicative error of $1 \pm \epsilon$
with probability $1 - \delta$ can be obtained in time polynomial
in $n$, $1/\epsilon$, and $1/\delta$. This follows from the fact
that for every graph $C_1^+(G) < n^3$
(see~\cite{AKLLR,CRSST,FeigeUpper}). The question of whether there
is a deterministic algorithm that approximates $C_1^+(G)$ within a
multiplicative error of $1 \pm \epsilon$ in time polynomial in $n$
and $1/\epsilon$ is open (see for example Chapter~8
in~\cite{AldousFill}). Here we provide a positive answer to this
question in the special case that the underlying graph is a tree.

\subsection{Related work}
\label{sec:related}

A survey of random walks in graphs is provided by
Lovasz~\cite{lovasz}.
 A related book in preparation
by Aldous
and Fill~\cite{AldousFill} is also available on the web. There are
also additional books that contain much information on random
walks in graphs, such as the recent book by Levin, Peres and
Wilmer~\cite{LevinPeresWilmer}. More information and appropriate
references for some of the well known claims that we make below
can be found in these references.

A random walk on a graph is a special case of a Markov chain, with
the vertices of the graph serving as the states of the Markov
chain, and the edges providing an implicit representation for the
transition probabilities. Some parameters of interest for random
walks are the
{\em  expected hitting time} (expected number of steps it takes to
get from one given vertex to another given vertex), the {\em
expected commute time} (expected number of steps to make a
round-trip between two given vertices) and the {\em expected cover
time} (the expected number of steps that it takes to visit all
vertices). By convention, throughout this paper, we omit the
qualifier {\it expected} when we deal with expectations of random
times, and write {\it hitting time} for the expected hitting time,
etc. When dealing with the actual random variables instead of
their expectations, we use the term {\it random hitting time},
etc.

The main result described in this manuscript refers to the {\em
cover and return time}, which requires walks to return to the
starting vertex after covering the graph. The hitting time and the
commute time can be computed in polynomial time (by solving a
system of linear equations). In particular, let us note here that
the commute time between any two adjacent vertices in a tree with
$n$ vertices is exactly $2(n-1)$. The cover time and cover and
return time can be computed in exponential time (again by solving
a system of linear equations, but the number of variables is
exponential in the size of the graph). It is not known whether
there is a polynomial time algorithm for computing either the
cover time or the cover and return time.

As noted earlier, there is a natural randomized algorithm that in
polynomial time estimates the cover time (or alternatively, the
cover and return time), up to some small error. The question of
whether in general the use of randomness helps (in a substantial
way) in the design of polynomial time algorithms (or in complexity
theoretic terms, is BPP=P?) has a natural counterpart in the
context of the cover time, namely, can a deterministic polynomial
time algorithm achieve as good an approximation of the cover time
as the randomized algorithm? This question has been studied in the
past, with moderate success.

Much of previous work dealt with the cover time from the worst
possible starting vertex in the graph. In this case, the maximum
hitting time serves as a lower bound on the cover time. Moreover,
as shown by Matthews~\cite{Matthews}, the cover time can exceed
the maximum hitting time by a factor of
at most $\ln n$. Hence the
hitting time (which is computable in deterministic polynomial
time) provides a $\ln n$ approximation to the cover time. An
extension of this approach leads to an algorithm with a better
approximation ratio of $O((\log\log n)^2)$~\cite{KKLV}. An
approach of upper bounding the cover time based on spanning trees
is presented in~\cite{CRSST}. In particular, when it is applied to
trees it implies that the cover and return time is at most
$2(n-1)^2$ (which is attained for a path with $n$ vertices), and
for general graphs it gives an upper bound of $n^3$ (which can be
improved to essentially $4n^3/27$ with more careful
analysis~\cite{FeigeUpper}). For some graphs, this approach based
on spanning trees gives a very good approximation of the cover
time.

When one seeks to estimate the cover time from a given vertex
(rather than from the worst possible vertex), the known bounds
deteriorate. The deterministic algorithms
known~\cite{RabinovichFeige,ChlamtacFeige} pay an extra $O(\log
n)$ factor in the approximation ratio compared to the
approximation ratios known from worst possible vertex. For the
special case of trees, some upper bounds are presented
in~\cite{FeigeCoupon}.

There are some special families of graphs for which the cover time
is known exactly (e.g., for paths, cycles and complete graphs), or
almost exactly (e.g., for balanced
trees~\cite{Aldoustree}
and for two and higher dimensional grids~\cite{DPRZ,AldousFill}).

\subsection{Our results}

Our main theorem is the following.

\begin{theorem}
\label{thm:main} There is a deterministic algorithm that given as
input a tree $T$ on $n$ vertices, a starting vertex $v$ and a
slackness parameter $\epsilon > 0$, outputs a value
$A(T,v,\epsilon)$ that approximates the cover and return time
$C_v^+(T)$ within a factor of $1 \pm \epsilon$. Namely,
$$(1 - \epsilon)A(T,v,\epsilon) \le C_v^+(T) \le (1 +
\epsilon)A(T,v,\epsilon).$$ The running time of the algorithm is
polynomial in $n/\epsilon$ (hence of the form $O(n^a/\epsilon^b)$
for some fixed constants $a > 0$ and $b > 0$).
\end{theorem}

Our proof is constructive in the sense that we actually describe
the algorithm. We remark (see Section~\ref{sec:Markov}) that the
algorithm extends almost without change to estimating the cover
and return time of arbitrary Markov chains on trees, though the
running time in this case is polynomial in the cover and return
time itself rather than in the number of states. (This distinction
was not necessary for simple random walks on trees because there
the cover time is bounded by a polynomial in the number of
vertices.) The algorithm also extends to the case when we are
given a set $S$ of vertices in the tree, and are required to
estimate the expected time by which a random walk on $T$ covers
the vertices of $S$ and returns to $v$. See Section~\ref{sec:set}.

The additive error in the approximation provided by
Theorem~\ref{thm:main} is at most $\epsilon C_v^+(T)$. As
$C_v^+(T) < 2n^2$ for every $n$-vertex tree,
see Section \ref{sec:related},
it follows that the
additive error is at most $\epsilon/2n^2$. The running time of the
algorithm remains polynomial in $n$ even if $\epsilon < 1/2n^2$,
and hence Theorem~\ref{thm:main} also provides approximations of
the cover and return time with arbitrarily small additive error.

The proof of Theorem~\ref{thm:main} as appears in
Section~\ref{sec-tda} applies to the cover and return time but not
to the cover time. It is possible to use the cover and return time
in conjunction with the hitting times from leaves of $T$ to $v$ in
order to obtain accurate estimates on the cover time. Further, the
algorithm in Theorem~\ref{thm:main} and its proof can be adapted
to handle also the cover time. Hence a statement similar to that
of Theorem~\ref{thm:main} (see Theorem~\ref{thm:cover}) applies
also to the cover time. We sketch the proof in
Section~\ref{sec:cover}.

\section{The deterministic algorithm}
\label{sec-tda}

Many computational problems that are difficult to solve on graphs
are easy (polynomial time solvable) on trees. The algorithmic
paradigm that is often used in these cases is dynamic programming.
We shall also use dynamic programming so as to approximate the
cover time on trees. The difficulty is that the cover time per se
is not a quantity that lends itself well to aggregation of
information. For example, consider a tree $T$ with root vertex $r$
connected to two vertices $r_1$ and $r_2$, which are root vertices
of subtrees $T_1$ and $T_2$. Even if one is given the complete
distribution function for the cover and return time of the
subtrees $T_1$ and $T_2$, it is not
immediately
clear (to the authors) how to
combine this information so as to obtain $C_r^+(T)$. To overcome
this difficulty, we extend an approach that was used by
Aldous~\cite{Aldoustree} for evaluating the cover time of
balanced trees.

Let $T$ be an arbitrary tree with vertices $v_1, \ldots v_n$ on
which we wish to estimate $C_1^+(T)$. For the sake of uniformity
of the notation, we shall introduce a new root vertex $r$ to the
tree connected only to $v_1$, thus obtaining a new tree that we
shall call $T_r$. The tree $T_r$ has $n+1$ vertices and $n$ edges.
For the sake of establishing notation, orient all edges away from
the root, and for every $1 \le i \le n$, let $e_i$ be the unique
edge whose endpoint is $v_i$. As a convention, we say that $e_i$
is {\em traversed} whenever the walk enters $v_i$ through $e_i$
(but not when the walk exits $v_i$ through $e_i$). Let $T_i$ be
the subtree rooted at $v_i$ (hence $T_1 = T$).
Now we define the key
quantity on which we shall employ dynamic programming.

\begin{definition}
\label{def:P} Using the notation introduced above, for $1 \le i
\le n$ and for $t \ge 1$, let $P_i(t)$ denote the probability that
a walk on $T_r$ that starts at $v_i$ visits all vertices of $T_i$
before edge $e_i$ is traversed $t$ times.
\end{definition}

As a simple example, if $v_i$ is a leaf of $T$, then $P_i(t) = 1$
for every $t$. This will serve as the base case that will start
off our dynamic programming. Our goal will be to compute $P_1(t)$
for all $t$. Using these values, we may consider $E(1) = \sum_{t}
(1 - P_1(t))$ which is equal to the expected number of times that
$e_1$ is traversed in a walk on $T_r$ that starts at $v_1$ and
covers $T_1$. (Technically, $E(1)$ is an infinite sum. However,
the sum converges since necessarily $E(1) < C_1^+(T_r)$, and
$C_1^+(T_r) \le 2n^2$.)

The following lemma shows the connection between the value of
$E(1)$ and the desired $C_1^+(T)$.

\begin{lemma}
\label{lem:translation} With notation as above, $C_1^+(T) =
2(n-1)E(1)$.
\end{lemma}

\begin{proof}
Let $C[v_1,r]$ denote the {\em commute time} between $v_1$ and $r$
in $T_r$ (the expected number of steps it takes a walk that starts
at $v_1$ to visit $r$ and return to $v_1$). As mentioned in
Section~\ref{sec:related},
$C[v_1,r] = 2n$. Observe that due to Wald's lemma,
$C_1^+(T_1) = E(1)C[v_1,r]$. (An intuitive way
to see the latter equality is by considering an extremely
long random walk on $T_r$ that starts at $v_1$, and covers $T_1$
many times. Break the walk  into
segments that correspond to
the walk covering
$T_1$ and returning to $v_1$.
During the first
$\ell$ such segments, with $\ell$ large, the ergodic theorem
implies that the number of commutes
to $r$ is close to $E(1)\ell$. Taking $\ell\to\infty$ then yields the
identity.) Moreover, observe that one can
relate the cover time of $T_1$ in $T_r$ to that in $T$ by
subtracting the steps along the edge $e_1$ (in both directions).
Linearity of expectation then implies that $C_1^+(T) = C_1^+(T_1)
- 2E(1)$. Putting everything together we deduce that $C_1^+(T) =
2(n-1)E(1)$.
\end{proof}

As noted above, to compute $E(1)$ from $P_1(t)$ involves an
infinite sum. To obtain a finite algorithm, we shall truncate the
sum when $t$ exceeds a sufficiently large value $N$. To keep the
presentation simple, we shall not attempt to optimize the value of
$N$ here (not for trees in general and not for any tree
specifically), but just note that $N$ can be chosen to be $O(n^2
\log 1/\epsilon)$, because it is not hard to show that
for some universal constant $c > 0$,
\begin{equation}
    \label{eq-ofer1}
    P_1(t) \ge 1 - e^{-ct/n^2} \,.
\end{equation}
Indeed, since $C_1^+(T_r)
\leq 2n^2$, the probability to cover $T_r$ within the first $4n^2$ traverses
of the edge $(r,v_1)$ is at least $1/2$, which implies the estimate
on $P_1(t)$. More
generally, taking $N$ as $O(n^2 \log (n + \frac{1}{\epsilon}))$ we
will be able for every vertex $v_i$ to consider the values of
$P_i(t)$ only for $t$ up to $N$, while still eventually achieving
a $(1 \pm \epsilon)$ multiplicative approximation for $C_1^+(T)$.

We now proceed to describe an exact dynamic programming procedure
in an idealized world in which computations can be done with
arbitrary precision and summations may include infinitely many
summands (though all sums do converge). Later we shall discuss how
the dynamic programming can be carried out in polynomial time with
only a small loss in the accuracy of the computations.

For every vertex $v_i$ we shall compute the infinite vector $P_i =
\{P_i(t)\}$ for all values of $t$. (Needless to say, in our actual
algorithm we shall truncate this vector an $t=N$.) As noted, for
every $v_i$ that is a leaf of $T_r$, this is the all~1 vector. For
every other vertex $v_i$, let $D_i$ denote the set of direct
descendants of $v_i$ (those vertices connected to $v_i$ by edges
other than $e_i$). Given that $T_r$ is a tree, it will always be
the case that if we have not yet computed $P_1$, then there is
some vertex $v_i$ for which $P_i$ has not yet been computed but
the vectors $P_j$ were already computed for all $v_j \in D_i$.
Hence we will compute $P_i$ for such a vertex $v_i$ and make
progress. The computation will involve quantities that shall be
defined next.

Fix a vertex $v_i$ of interest. To simplify notation, let $d =
|D_i|$ be the number of direct descendants of $v_i$. Rename them
as $u_1, \ldots, u_d$.

\begin{definition}
\label{def:Q} For vertex $v_i$, $t \ge 1$, $t_1 \ge 1, \ldots, t_d
\ge 1$, define $Q_i(t_1, \ldots, t_d;t)$ to be the probability
that in a walk on $T_r$ that starts at $v_i$, each edge
$(v_i,u_j)$ is traversed exactly $t_j$ times (in the direction
into $u_j$) before the edge $e_i$ is traversed $t$ times.
\end{definition}

Observe (though we shall not need to use this fact) that for two
vertices $v_i$ and $v_j$ with the same number of descendants, the
functions $Q_i$ and $Q_j$ are identical.

We now have a recursive formula for $P_i(t)$ in terms of the
vectors $P_j$ for the descendants $v_j \in D_i$. (So as to keep
notation simple, we use in this formula the convention that $P_i$
refers to $v_i$, but $P_j$ refers to $u_j$ rather than $v_j$.)

\begin{equation}
\label{eq:exact} P_i(t) = \sum_{t_1 \ge 1, \ldots, t_d \ge 1}
Q_i(t_1, \ldots, t_d;t) \prod_{j=1}^d P_j(t_j)
\end{equation}

Let us explain Equation~(\ref{eq:exact}). We wish to compute the
probability that $T_i$ is covered before $e_i$ is traversed $t$
times. In order to cover $T_i$, each vertex of $u_j \in D_i$ must
be visited at least once, and the subtree $T_j$ rooted at $u_j$
needs to be covered. Once we fix the stopping condition of the
edge $e_i$ being traversed $t$ times, the distribution of the
number of visits (from their parents)
to the descendants $u_j \in D_i$ is given by the
function $Q_i$. Subtree $T_j$ needs to be covered by the time the
edge $(v_i,u_j)$ is traversed $t_j$ times, one of which is the
first entry to $u_j$, and hence the term $P_j(t_j)$ gives the
probability of $T_j$ being covered. We can take the product of the
terms $P_j(t_j)$, because the walks within different subtrees are
independent.

Using Equation~\ref{eq:exact} and the fact that the vectors $P_i$
are known for all leaves, we get an inductive definition for
$P_1$, and then Lemma~\ref{lem:translation} can be used to compute
$C_1^+(T)$. However, there are several obstacles to obtaining a
polynomial time algorithm. We list these obstacles, and then
explain how to overcome them, paying only a multiplicative factor
of $(1 \pm \epsilon)$ in the accuracy of the computation.

\begin{enumerate}
\item {\bf Range of summation.}
Each variable $t_j$ ranges over infinitely many values. As
explained earlier, this will be handled by limiting the range
between~1 and $N$ for sufficiently large $N$.
\item {\bf Combinatorial explosion.}
Even if the range of the summation of each variable is limited to
$N$, the number of terms in the summation is $N^d$. Since $d$ need
not be bounded by a constant (the tree may have vertices of
arbitrarily large degrees), this number will not be polynomial in
$n$. We shall refine the dynamic programming approach so as to
overcome this obstacle.
\item {\bf Finite precision.}
Computation cannot be performed with infinite precision. We
shall either need to show that the numbers involved can always be
represented using polynomially many bits, or round some of the
numbers and account for the error introduced by the rounding.
\end{enumerate}

It would be more convenient for us to first deal with the second
obstacle, and only later with the other obstacles.

\subsection{Avoiding the combinatorial explosion}
\label{sec:2.1}

For every vertex $v_i$, if $|D_i| > 2$, construct an arbitrary
binary tree $B_i$ (each internal node has two children) with $d =
|D_i|$ leaves, placing $v_i$ at the root and $u_1, \ldots, u_{d}$
at the leaves. There are $d-2$ internal nodes (in addition to the
root) that we shall name as $b^i_1, \ldots, b^i_{d-2}$.
For simplicity of notation, let us fix
the structure of the tree to be a path $v_i, b^i_1, \ldots
b^i_{d}$, with $u_1$ connected to $v_i$, $u_d$ connected to
$b^i_{d-2}$, and $u_j$ for $1 < j < d$ connected to $b^i_{j-1}$.

The random walk on $T_r$ can be simulated as follows. Whenever the
walk on $T_r$ reaches $v_i$, with probability $1/(d+1)$ it takes
the edge $e_i$, and with probability $d/(d+1)$ it goes to one of
the children, chosen uniformly at random. This random choice of
child is simulated by a walk on the tree $B_i$. Conditioned on
having decided not to take the edge $e_i$, at every internal node
of the tree $B_i$ (including the root), choose one of the two
children with probability proportional to the number of leaves of
$B_i$ that are descendants of the child. For example, at internal
node $b^i_k$ with $k < d-2$, go to leaf $v_{k+1}$ with probability
$1/(d-k)$ and to internal node $b^i_{k+1}$ with probability
$(d-k-1)/(d-k)$. It can readily be seen that each leaf is reached
with the same probability. Being at a leaf $u_j$ in $T_r$ and
deciding to take the edge $(u_j,v_i)$ is simulated in $T_B$ by
taking the path $u_j$ to $v_i$ in the tree $B_i$.

For the simulated random walk, every vertex has only two children.
This is the key to avoiding the combinatorial explosion. Observe
that building such trees $B_i$ for all vertices $v_i$, we change
$T_r$ into a tree $T_B$ which is a subtree of  the  binary tree.
Every leaf of $T_B$ is a leaf of
$T_r$, and so it follows that the total number of vertices in
$T_B$ is at most $2n$.

The tree $T_B$ is still rooted at $r$ like $T_r$, and $r$ has
degree~1 also in $T_B$. Except for $r$, $T_B$ has two types of
vertices: those which were original vertices of $T$ (and were
denoted by $v_i$), and those that were added by the subtrees $B_i$
(and were denoted by $b^i_k$). For uniformity of notation, we use
$w_i$ to denote vertices of $T_B$, regardless of the origin of the
vertex.  However, we associate with each vertex $w_i$ a weight
$W_i$. The weight of each of the original vertices of $T_r$ is~1.
The weight of a vertex $b^i_k$ of $B_i$ is always
greater than~1, and equal to the number of leaves of $B_i$ in the
subtree of $B_i$ rooted at $b^i_k$. (With the notation that we
used above, it turns out that this weight is equal to $d-k$.) As
in the case of $T_r$, we now use $T_i$ to denote the subtree of
$T_B$ rooted at $w_i$.

Recall that a walk on a graph is a sequence of vertices (that respects the
adjacency structure of the graph). We now define a (random) walk
$\{S_n\}$ on $T_B$, as follows.

\begin{enumerate}
\item At $r$, move to its unique neighbor $w_1$.
\item Let the walk be at a vertex $w_i$ with $W_i = 1$ (hence, an
original vertex of $T$). Let $w_p$ be its parent node.
\begin{enumerate}
\item If $w_i$ is a leaf, move to its parent vertex $w_p$.
\item If $w_i$ has only one child, move to this child with
probability $1/2$ and to $w_p$ with probability $1/2$.
\item Otherwise, $w_i$ must have exactly two children, one of them (say $w_l$) of weight~1 and the other
(say $w_r$) of weight $W_r \ge 1$. Move to $w_p$ with probability
$1/(2 + W_r)$, to $w_l$ with probability $1/(2 + W_r)$, and to
$w_r$ with probability $W_r/(2 + W_r)$.
\end{enumerate}
\item Let the walk be at a vertex $w_i$ with $W_i > 1$ (hence a
vertex that was introduced through some subtree $B$). Let $w_p$ be
its parent node, and $w_l$ and $w_r$ be its two children. At least
one of these children is an original vertex of $T_r$, hence we
assume without loss of generality that $W_l = 1$.
\begin{enumerate}
\item If $w_i$ was last entered from one of its children, move to
$w_p$.
\item If $w_i$ was last entered from $w_p$, move to $w_l$ with
probability $1/(1 + W_r)$ and to $w_r$ with probability
$W_r/(1+W_r)$.
\end{enumerate}
\end{enumerate}
Note that the random walk thus defined is not Markovian, while
the process $\{(S_{n-1},S_n)\}_{n\geq 1}$ is Markovian.

So far, we have defined two random walk
processes, one on $T_r$ and one on $T_B$. For a walk on $T_B$,
we now define the {\em
projection} of the walk to be the subsequence of vertices that
includes only the original vertices of $T_r$ (removing the
vertices introduced by the subtrees $B$ from the sequence). Random
walks on $T_B$ simulate random walks on $T_r$ in the sense that
the projection of a random walk on $T_B$ is precisely a random
walk on $T_r$.

Definition~\ref{def:P} applies with minor changes to walks on
$T_B$. We present the revised definition.

\begin{definition}
\label{def:P2} For vertex $w_i$ in $T_B$, with parent vertex
denoted by $w_p$, and for $t \ge 1$, define $P_i(t)$ to be the
probability of the following event:
\begin{enumerate}
\item If
$W_i = 1$, the event is that a walk on $T_B$ that starts at $w_i$
visits all vertices of $T_i$ before traversing the edge
$(w_p,w_i)$ $t$ times.
\item If $W_i > 1$, the event is that a walk on $T_B$ that just entered
$w_i$ from $w_p$ visits all vertices of $T_i$ before traversing
the edge $(w_p,w_i)$ $t$ additional times.
\end{enumerate}
\end{definition}

Likewise, Definition~\ref{def:Q} needs to be modified so as to
account for the existence of different types of vertices in $T_B$.

\begin{definition}
\label{def:Q2} For vertex $w_i \not= r$ in $T_B$, let $w_p$ denote
its parent vertex and let $w_l$ and $w_r$ denote its two children
(or only $w_l$ if $w_i$ has one child). For $t \ge 1$, $t_l \ge
1$, $t_r \ge 1$, define $Q_i(t_l,t_r;t)$ (or $Q_i(t_l;t)$ if $w_i$
has only one child) to be the probability of the following event:
\begin{enumerate}
\item
If $W_i = 1$, then the event is that in a walk on $T_B$ that
starts at $w_i$, edge $(w_i,w_l)$ is traversed exactly $t_l$ times
and edge $(w_i,w_r)$ is traversed exactly $t_r$ times  before the
edge $(w_p,w_i)$ is traversed $t$ times. (If $w_i$ has only one
child, then remove the condition on $t_r$.)
\item
If $W_i > 1$, then the event is that in a walk on $T_B$ that just
entered $w_i$ from $w_p$, edge $(w_i,w_l)$ is traversed exactly
$t_l$ times and edge $(w_i,w_r)$ is traversed exactly $t_r$ times
before the edge $(w_p,w_i)$ is traversed $t$ additional times.
\end{enumerate}
\end{definition}

Armed with the new definitions for $P_i$ and $Q_i$,
Equation~(\ref{eq:exact}) when applied to $T_B$ simplifies to:
\begin{equation}
\label{eq:degree2} P_i(t) = \sum_{t_1 \ge 1, t_2 \ge 1} Q_i(t_1,
t_d;t) P_1(t_1)\cdot P_2(t_2)
\end{equation}

Inductively applying Equation~(\ref{eq:degree2}) in $T_B$ we obtain
the vector $P_1$ in $T_B$, which is equal to the vector $P_1$ in
$T_r$.

\subsection{Limiting the range of summation}
\label{sec:2.2}

For some sufficiently large value of $N$ (to be determined later),
we shall truncate all vectors $P_i$ after $N$ entries, implicitly
assuming that $P_i(t) = 1$ for all $i \ge N$. Hence we shall set
$P_i(N) = 1$, regardless of its true value or computed value. For
vertices $w_i$ that are not leaves, this certainly introduces an
error. Moreover, this error propagates and amplifies through our
use of Equation~(\ref{eq:degree2}).

We shall modify Definition~\ref{def:Q2} to reflect the fact that
we no longer distinguish between different values of $t_j$ that
are larger than $N$.

\begin{definition}
\label{def:revisedQ} For vertex $w_i \not= r$ in $T_B$, let $w_p$
denote its parent vertex and let $w_l$ and $w_r$ denote its two
children (or only $w_l$ if $w_i$ has one child). For $1 \le t \le
N$, $1 \le t_l \le N$, $1 \le t_r \le N$, define $Q_i(t_l,t_r;t)$
(or $Q_i(t_l;t)$ if $w_i$ has only one child) to be the
probability of the following event:
\begin{enumerate}
\item
If $W_i = 1$, then the event is that in a walk on $T_B$ that
starts at $w_i$, edge $(w_i,w_l)$ is traversed exactly $t_l$ times
(and at least $t_l$ times in the special case that $t_l = N$) and
edge $(w_i,w_r)$ is traversed exactly $t_r$ times  (and at least
$t_r$ times in the special case that $t_r = N$) before the edge
$(w_p,w_i)$ is traversed $t$ times. (If $w_i$ has only one child,
then remove the condition on $t_r$.)
\item
If $W_i > 1$, then the event is that in a walk on $T_B$ that just
entered $w_i$ from $w_p$, edge $(w_i,w_l)$ is traversed exactly
$t_l$ times (and at least $t_l$ times in the special case that
$t_l = N$) and edge $(w_i,w_r)$ is traversed exactly $t_r$ times
(and at least $t_r$ times in the special case that $t_r = N$)
before the edge $(w_p,w_i)$ is traversed $t$ additional times.
\end{enumerate}
\end{definition}

We can now modify our recursive formula to have only finitely many
terms. It no longer computes the true value of $P_i(t)$, so we
shall call the quantity that it computes $P^1_i(t)$. The function
$Q$ to be used in this formula is the one from
Definition~\ref{def:revisedQ}. $P^1_i(N)$ is not computed by this
formula, but instead set to~1.

\begin{equation}
\label{eq:finite} P^1_i(t) = \sum_{1 \le t_l \le N, 1 \le t_r \le
N} Q_i(t_l, t_r;t) P^1_l(t_l)P^1_r(t_r)
\end{equation}

This completes the description of how we limit the range of
summation to be finite. We now analyze the effect of this
approximation. For a given choice of $N$, let $\delta > 0$ be such
that for every $i$, $(1 + \delta)P_i(N) \ge 1$. For concreteness, take
\begin{equation}
    \label{eq-ofer2}
    \delta= 2(1-P_i(N))\leq 2 e^{-cN/n^2}\,,
\end{equation}
see
(\ref{eq-ofer1}). We shall express
the relative error in the approximation as a function of $N$ and
$\delta$, and thereafter choose $N$ such that together with the
implied $\delta$, the relative error is smaller than $\epsilon$.

At the leaves of $T_B$ there is no error in the respective vector
$P_i$. At a vertex $w_i$ whose two children (or single child, if
$w_i$ has only one child) are leaves, a  multiplicative error of
at most $(1 + \delta)$ is introduced because $P^1_i(N)$ is rounded
to~1, even though its true value may have been $1/(1 + \delta)$.
Consider now some other arbitrary vertex $w_i$, let $w_l$ and
$w_r$ be its children, and let $(1 + \delta_l)$ and $(1 +
\delta_r)$ be upper bounds on the multiplicative errors in any of
the entries of the vectors $P^1_l$ and $P^1_r$. Then by inspection
of Equation~(\ref{eq:finite}), the multiplicative error in any entry
of $P^1_i$ is at most $(1 + \delta_l)(1 + \delta_r)$.  Since $T_B$
has at most $2n$ vertices, it follows that the multiplicative
error at entries of $P^1_1$ (compared to the true entries of
$P_1$) is at most $(1 + \delta)^{2n}$.

Recall that we needed the vector $P_1$ so as to compute the
expectation $E(1) = \sum_{t \ge 1} (1 - P_1(t))$. Instead we now
compute at approximation $E^1(1) = \sum_{1 \le t \le N} (1 -
P^1_1(t))$. Hence our total error in this computation is:

$$E(1) - E^1(1) = \sum_{1 \le t \le N} (P^1_1(t) - P_1(t)) +
\sum_{t > N} (1 - P_1(t))$$

The first of these summations is at most $N((1 + \delta)^{2n} -
1)$. If $\delta \ll \frac{1}{2n}$ then this value is approximated
well by $2nN\delta$. In the second of these summations, each term
is of value at most $\delta$. Moreover, for every $t$, $(1 -
P_1(t+N)) \le \delta(1 - P_1(t))$. Hence if $\delta < 1/2$ then
the second summation can be upper bounded by a geometric series of
sum $2N\delta$. Hence the total additive error is at most
$2(n+1)N\delta$, and we wish it to be smaller than $\epsilon
E(1)$. It is not hard to show that in every tree $E(1) \ge 1$ (in
fact, in every tree $E(1)$ is essentially the cover time divided
by $2n$, and the cover time of a graph is $\Omega(n\log n)$), and
hence we shall simplify the desired inequality to $nN\delta \le
\epsilon$. This requires choosing $N$ such that $\delta \le
\frac{\epsilon}{nN}$. With the choice of $\delta$ in (\ref{eq-ofer2}),
a value of $N =
cn^2\log(n + \frac{1}{\epsilon})$ for a sufficiently large constant
$c$ would suffice for all trees.

We remark that in this paper we just give a sufficient value of
$N$. Much lower values of $N$ will also work for special families
of trees (essentially, a factor of $n$ can be replaced by their
cover time divided by $n$), and moreover, we need not use the same
value of $N$ for all vertices of $T_B$ (in particular, for the
leaves we may take $N = 1$). These kind of optimizations are
omitted from this paper.

\subsection{Computation with finite precision}
\label{sec:2.3}

Having established the value of $N$ for which $P^1_1$ is a
sufficiently close approximation for $P_1$, it remains to verify
that $P^1_1$ can indeed be computed in polynomial time. For this,
we need to be able to compute the values $Q_i(t_l,t_r;t)$. Let us
first observe that the role of vertex $w_i$ in the value of this
expression is only in determining the weight of $w_r$ (the weight
of $w_l$ is always~1). Definition~\ref{def:revisedQ} offers
several cases for the definition of $Q_i$, and we shall address
only some of them here. The other cases are handled similarly.

Let us compute $Q_i(t_l,t_r;t)$ when $W_i = 1$, $t_l \not= N$ and
$t_r \not= N$. Whenever the walk is at $w_i$, it has probability
$p_p = 1/(2 + W_r)$ to go to $w_p$, probability $p_l = 1/(2 +
W_r)$ to go to $w_l$, and probability $p_r = W_r/(2 + W_r)$ to go
to $w_r$. The probability of exactly $t_l$ visits to $w_l$ and
exactly $t_r$ visits to $w_r$ prior to $t$ visits to $w_p$ is
exactly $${t + t_l + t_r - 1 \choose t_l}{t + t_r - 1 \choose
t_r}(p_p)^t(p_l)^{t_l}(p_r)^{t_r}\,.$$ The upper bound of $N$ implies
that both the numerator of this expression and the denominator are
numbers that can be expressed by $O(N(\log N + \log n))$ bits.

If $t_r = N$ then $Q_i(t_l,N;t)$ will be computed differently.
First, ignoring moves into $w_r$ (as if $w_i$ has only one child),
compute $Q_i(t_l;t)$. Then subtract $\sum_{0 \le t_r \le
N-1}Q_i(t_l,t_r;t)$ to get the desired result. Observe that for
the final answer one can use a common denominator $(N!)^2(2 +
W_r)^N$, and hence still expressible in a polynomial number of
bits.

A similar argument applies to the computation of
$Q_i(t_l,t_r;t)$ when $W_i > 1$, $t_l \not= N$ and
$t_r \not= N$. In this case,
the probability of exactly $t_l$ visits to $w_l$ and
exactly $t_r$ visits to $w_r$ prior to $t$ additional
visits to $w_p$
vanishes unless $t=t_l+t_r$, in which case it equals
 $${t   \choose t_l}
\frac{W_r^{t_r}}{(1+W_r)^{t}}\,,$$
where $W_r$ is the weight of the right descendent of $W_i$.

Following the computation in Equation~(\ref{eq:finite}), and
thereafter applying it to all vertices of $T_B$, one sees that one
can obtain a rational number with denominator $(N!n!)^{O(n)}$, and
likewise with a numerator expressible by polynomially many bits.
Hence in principle, all computations can be performed exactly in
polynomial time, though they would be very tedious.

A more practical approach is to round the numbers to numbers of
shorter representations. Clearly, this can be done while
maintaining the relative error in the range $(1 \pm \epsilon)$,
but we omit concrete suggestions of how to do this.

\section{Extensions}
We present in this section several extensions of Theorem
\ref{thm:main}.

\subsection{Arbitrary Markov chains on trees}
\label{sec:Markov} Consider a Markov chain $\{S_t\}$ with state
space the vertices of a (finite) tree $T$, where transitions are
allowed only between neighbors. Because of the tree structure, the
Markov chain is reversible, and hence there exist conductances
${\cal C}_{\{u,v\}}$ (with $u,v$ neighboring vertices in the tree)
such that the transition probability from $u$ to $v$ equals ${\cal
C}_{\{u,v\}}/\sum_{w: w\sim u} {\cal C}_{\{u,w\}}$. Let ${\cal
R}_{\{u,v\}} = 1/{\cal C}_{\{u,v\}}$ denote the respective
resistance between neighboring vertices in the tree, and assume
first that all resistances are integer valued. Consider the  tree
$T'$ in which each edge $\{u,v\}$  is replaced by a chain of
length ${\cal R}_{\{u,v\}}$, and a simple random walk $\{RW_t\}$
on $T'$. Thus, each vertex of $T$ corresponds to a vertex of $T'$.
Further, the random walk $\{RW_t\}$ induces a Markov process on
$T$, and the transition probabilities of the latter coincide with
those of $\{S_t\}$. In particular, the quantity $E(1)$
corresponding to $T$ is identical to that corresponding to $T'$,
and can be computed accurately by Theorem \ref{thm:main}. For the
tree $T$, we have $C[v_1,r]=2\sum {\cal R}_{\{u,v\}}$, and
$C_1^+(T)=(C[v_1,r]-2) E(1)$ by an adaptation of Lemma
\ref{lem:translation}. We conclude from these facts that the cover
and return time can be computed by the algorithm of Theorem
\ref{thm:main}, with the running time polynomial in the cover and
return time itself rather than in the number of states. It is
straightforward to approximate the above in case the resistances
are not integer-valued.

\subsection{Covering a specified set of vertices}
\label{sec:set}
Let $T'$ be a subtree of $T$, rooted at $v_1$.
The algorithm of Theorem \ref{thm:main}
applies equally well to the
evaluation of the cover and return time of $T'$ by a random walk
on $T$, denoted $C_1^+(T';T)$, as follows. The quantity
$E(1)$ for $T'$, denoted $E_{T'}(1)$,
can be computed  by the algorithm (applied to $T'$).
We then have (again, by an adaptation of Lemma \ref{lem:translation})
that $C_1^+(T';T)=2(n-1)E_{T'}(1)$.

\subsection{Computing the cover time}
\label{sec:cover}

Fix a tree $T$ and a starting vertex $v$. Given a vertex $u$, let
$P_{last}[u]$ denote the probability that for a random walk on $T$
that starts at $v$, the last vertex to be visited is $u$. Clearly,
$P_{last}[u] \not= 0$ iff $u$ is a leaf of $T$ (different than
$v$). Let $H[u,v]$ denote the expected hitting time in $T$ from
$u$ to $v$. Then the cover time satisfies:

$$C_v(T) = C_v^+(T) - \sum_u P_{last}[u] H[u,v]$$

Recall that for every vertex $u$, $H[u,v]$ can be computed exactly
in polynomial time (moreover, the known algorithms compute
$H[u,v]$ for all $u$ simultaneously, though this fact is not
needed here), and that $C_v^+(T)$ can be computed with arbitrary
small additive error. It follows that it suffices to estimate the
quantities $P_{last}[u]$ with sufficiently high precision in order
to obtain an accurate estimate of the cover (without return) time.

The latter task can be performed in a way similar to that
described in Theorem \ref{thm:main}. We sketch the steps, assuming
a reduction to a binary tree $T_B$ has already been performed as
in Section \ref{sec-tda}.

We begin with a definition.

\begin{definition}
For a vertex $v_i$ with $u \in T_i$ and for $t \ge 1$, let $A_i(t)$
denote the probability that a walk on $T_B$ that starts at $v_i$
satisfies the following conditions.

\begin{itemize}

\item It does not visit $u$ before edge $e_i$ is traversed $t-1$
times.
\item It does visit $u$ by the time edge $e_i$ is traversed $t$
times.
\item $u$ is the last vertex from $T_i$ to be visited.

\end{itemize}
\end{definition}

Clearly, $P_{last}[u] = \sum_t A_v(t)$, where $v$ is the starting
vertex of the walk and $u\in T_v$.

The following definition is similar to Definition~\ref{def:P}.
It will be used later in situations where $u\not\in T_i$.

\begin{definition}
For a vertex $v_i$
and for $t \ge 1$, let
$P_i(t)$ denote the probability that a walk on $T_B$ that starts
at $v_i$ visits all vertices of $T_i$  before edge $e_i$ is
traversed $t$ times.
\end{definition}

Note that we have already seen in Section~\ref{sec-tda} how all
$P_i(t)$ can be computed. We now explain how this can be used in
order to compute all $A_i(t)$.

\begin{definition}
    \label{def:3.3}
For a vertex $v_i$ with two children ($v_l$ and $v_r$), with $u \in
T_{v_l}$, $t \ge 1$, $t_l \ge 1$, $t_r \ge 1$, let
$R_i(t_l,t_r;t)$ denote the probability that a walk on $T_B$ that
starts at $v_i$ satisfies the following conditions.
\begin{itemize}
\item By the time edge $e_i$ is traversed $t-1$ times, the edge
$e_l$ is traversed at most $t_l - 1$ times.
\item By the time edge $e_i$ is traversed $t$ times, the edge
$e_l$ is traversed at least $t_l$ times.
\item By the time edge $e_l$ is traversed $t_l$ times, the edge
$e_r$ is traversed exactly $t_r$ times.
\end{itemize}
\end{definition}

The function $R_i(t_l,t_r;t)$ can be computed efficiently in a way
similar to that described in Section~\ref{sec:2.3}. The details
are tedious and are omitted. Now, with $u$ as in Definition
\ref{def:3.3},
$A_i(t)$ can be computed using
the following recursive formula:

$$A_i(t) = \sum_{t_l \ge 1, t_r \ge 1}
R_i(t_l,t_r;t)A_l(t_l)P_r(t_r)$$

The truncation of the sum to a finite sum can
performed as in Section \ref{sec:2.2}, with a similar computational cost.

The outline above (together with additional technical details
which are omitted) implies the following theorem.

\begin{theorem}
\label{thm:cover} There is a deterministic algorithm that given a
tree $T$ on $n$ vertices, a starting vertex $v$ and a slackness
parameter $\epsilon > 0$, outputs a value $A(T,v,\epsilon)$ that
approximates the cover time $C_v(T)$ within a factor of $1 \pm
\epsilon$. Namely,
$$(1 - \epsilon)A(T,v,\epsilon) \le C_v(T) \le (1 +
\epsilon)A(T,v,\epsilon).$$ The running time of the algorithm is
polynomial in $n/\epsilon$.
\end{theorem}

\subsection*{Acknowledgements}

The work of the authors is supported in part by The Israel Science
Foundation (grants No. 873/08 and 938/07, respectively).


\end{document}